\global\def\draftcontrol{0}

   \def\versionno{ alpha prime tauP and kappa -- draft   }

\catcode`\@=11

\expandafter\ifx\csname draftcontrol\endcsname\relax\global\def\draftcontrol{0}
\fi

{\count255=\time\divide\count255 by 60
\xdef\hourmin{\number\count255}
\multiply\count255 by-60\advance\count255 by\time
\xdef\hourmin{\hourmin:\ifnum\count255<10 0\fi\the\count255}}
\def\draftdate{\number\month/\number\day/\number\year\ \ \ \hourmin }

\newcommand\makepapertitle{\par
  \begingroup
    \renewcommand\thefootnote{\@fnsymbol\c@footnote}%
    \def\@makefnmark{\rlap{\@textsuperscript{\normalfont\@thefnmark}}}%
    \long\def\@makefntext##1{\parindent 1em\noindent
            \hb@xt@1.8em{%
                \hss\@textsuperscript{\normalfont\@thefnmark}}##1}%
     \newpage
     \global\@topnum\z@   
     \@makepapertitle
     \thispagestyle{empty}\@thanks
  \endgroup
  \setcounter{footnote}{0}%
  \global\let\thanks\relax
  \global\let\makepapertitle\relax
  \global\let\@makepapertitle\relax
  \global\let\@thanks\@empty
  \global\let\@author\@empty
  \global\let\@date\@empty
  \global\let\@title\@empty
  \global\let\title\relax
  \global\let\author\relax
  \global\let\date\relax
  \global\let\and\relax
  \def\version{\let\version\@version\@gobble}
}
\def\@makepapertitle{%
  \newpage
   \ifnum\draftcontrol=1 {}
   \version\versionno
   \vskip 3em%
   \else
   \hfill\hbox to 3cm {\parbox{4cm}{\@pubnum}\hss}%
   \vskip 3em%
   \fi
   \begin{center}%
   \let \footnote \thanks
     {\LARGE {\@title}}%
     \vskip 1.5em%
     {\normalsize
       \lineskip .5em%
       \begin{tabular}[t]{c}%
         \@author
       \end{tabular}\par}%
     \vskip 1.5em%
     {\@bstract}%
     \end{center}%
     \vskip 1.5em
     \@date%
   \par
}

\gdef\@pubnum{}
\def\pubnum#1{%
  \gdef\@pubnum{#1}}

\gdef\@bstract{}
\def\Abstract#1{%
  \gdef\@bstract{%
   \parbox{\textwidth-0pc}{%
   \centerline{\bf Abstract}\penalty1000%
\kern.2cm%
\noindent
\renewcommand\baselinestretch{1.0}%
{#1}}}
}

\def\ps@paper{\let\@mkboth\@gobbletwo%
     \ifnum\draftcontrol=1
	\def\@oddfoot{\hbox to \textwidth{\tiny \versionno \hfil\tiny\draftdate}%
	\hskip -\textwidth \hbox to \textwidth{\hfil\rm\thepage\hfil}}%
     \else\def\@oddfoot{\hbox to \textwidth{\hfil\rm\thepage\hfil}}
     \fi
     \let\@evenfoot\@oddfoot
}

\def\body{\clearpage
          \pagestyle{paper}
	}

\def\@version#1{\ifnum\draftcontrol=1
\typeout{}\typeout{#1}\typeout{}
\vskip3mm\centerline{\hbox{\fbox{\normalsize{\tt DRAFT -- #1 -- }
                   {\draftdate}}}}\vskip3mm
\fi}
\let\version\@version
\long\def\eqlabel#1{\ifnum\draftcontrol=1
                    \tag@false  
                    \tag*{(\theequation) \hbox to -0.2cm{\hspace{0cm}\small{#1}\hss}}
                    \refstepcounter{equation}
                    \edef\@currentlabel{\theequation}
                    \ltx@label{#1}          
                    \else
                    \label{#1}
                    \fi
                    }
\let\st@bibitem\@bibitem
\let\st@lbibitem\@lbibitem
\ifnum\draftcontrol=1
  \def\@bibitem#1{%
    \st@bibitem{#1}\a@@label{#1}\ignorespaces}
  \def\@lbibitem[#1]#2{%
    \st@lbibitem[#1]{#2}\a@@label{#2}\ignorespaces}
  \def\a@@label#1{%
    \gdef\a@lab{\smash{\normalfont\small#1}}
    \ifvmode
      \if@inlabel
        \global\setbox\@labels\hbox{%
          \llap{\a@lab\let\a@lab\relax
                \kern\@totalleftmargin\kern\marginparsep}%
          \box\@labels}%
      \fi
    \fi}
\fi

\documentclass[12pt,letterpaper]{article}

\usepackage{amsmath,amssymb,array,calc,rotating,epsfig,psfrag,bm}
\usepackage[nosort]{cite}

\ifnum\draftcontrol=1
\fi

\renewcommand\baselinestretch{1.25}
\renewcommand\section{\@startsection {section}{1}{\z@}%
                                   {-3.5ex \@plus -1ex \@minus -.2ex}%
                                   {2.3ex \@plus.2ex}%
                                   {\normalfont\large\bfseries}}
\renewcommand\subsection{\@startsection{subsection}{2}{\z@}%
                                   {-3.25ex\@plus -1ex \@minus -.2ex}%
                                   {1.5ex \@plus .2ex}%
                                   {\normalfont\normalsize\bfseries}}
\renewcommand\subsubsection{\@startsection{subsubsection}{3}{\z@}%
                                   {-3.25ex\@plus -1ex \@minus -.2ex}%
                                   {1.5ex \@plus .2ex}%
                                   {\normalfont\normalsize\it}}
\renewcommand\paragraph{\@startsection{paragraph}{4}{\z@}%
                                   {-3.25ex\@plus -1ex \@minus -.2ex}%
                                   {1.5ex \@plus .2ex}%
                                   {\normalfont\normalsize\bf}}

\numberwithin{equation}{section}

\def\ie{{\it i.e.}}

\def\revise#1       {\raisebox{-0em}{\rule{3pt}{1em}}%
                     \marginpar{\raisebox{.5em}{\vrule width3pt\
                     \vrule width0pt height 0pt depth0.5em
                     \hbox to 0cm{\hspace{0cm}{%
                     \parbox[t]{4em}{\raggedright\footnotesize{#1}}}\hss}}}}

\def\calc         {{\cal C}}

\def\calf         {{\cal F}}

\def\cali         {{\cal I}}

\def\calk         {{\cal K}}

\def\caln         {{\cal N}}
\def\calo         {{\cal O}}

\def\calt         {{\cal T}}

\def\sqr#1#2{{\vcenter{\vbox{\hrule height.#2pt
 \hbox{\vrule width.#2pt height#1pt \kern#1pt
 \vrule width.#2pt}\hrule height.#2pt}}}}

\def\a{\alpha}

\def\g{\gamma}

\newcommand{\qq}{\mathfrak{q}}

\newcommand{\ww}{\mathfrak{w}}

\def\p{\varphi}

\newcommand{\x}{\bm{x}}
\newcommand{\ka}{\bm{k}}
\def\dilog{{\rm dilog}}
\def\hw{\hat{\mathfrak{w}}}
\def\hq{\hat{\mathfrak{q}}}

\catcode`\@=12

\begin{document}


\title{Relaxation time of a CFT plasma at finite coupling}

\pubnum{%
UWO-TH-08/9\\
DAMTP-2008-49
}
\date{June 2008}

\author{
Alex Buchel$ ^{1,2}$ and Miguel Paulos$ ^3$\\[0.4cm]
\it $ ^1$Department of Applied Mathematics\\
\it University of Western Ontario\\
\it London, Ontario N6A 5B7, Canada\\[0.2cm]
\it $ ^2$Perimeter Institute for Theoretical Physics\\
\it Waterloo, Ontario N2J 2W9, Canada\\[0.2cm]
\it $ ^3$Department of Applied Mathematics and \\
\it Theoretical Physics, Cambridge CB3 0WA, U.K.
    }

\Abstract{
Following recent formulation of second order relativistic viscous
hydrodynamics for conformal fluids, we compute finite coupling
corrections to the relaxation time of $\caln=4$ supersymmetric
Yang-Mills plasma. The result is expected to be universal for any
strongly coupled conformal gauge theory plasma in four dimensions.
}


\makepapertitle

\body

\version\versionno

\section{Introduction and Summary}
To a large extend motivated and guided by the gauge theory/string theory correspondence of Maldacena 
\cite{m9711,m2,ss}, Baier {\it et.al} \cite{s3} and Bhattacharyya {\it et.al} \cite{m4}  recently formulated second order relativistic viscous 
hydrodynamics of conformal fluids. It was found that Mueller-Israel-Stewart (MIS) theory \cite{m,is},
often used in hydrodynamic simulations of strongly coupled quark-gluon plasma (sQCD) produced in heavy ion collisions 
at RHIC, does not properly account for all second order viscous corrections. The framework of this new viscous hydrodynamics 
was applied to bulk physics analysis at RHIC in \cite{lr}. 

In this paper we continue the program of computing the transport coefficients of strongly coupled gauge theories from the 
dual string theory holographic description. The motivation to do so, even though we lack a tractable string theory 
dual to QCD, is the amazing regularity of transport coefficients for a large class of gauge theory plasma at strong 't Hooft
coupling \cite{u1,u2,u3,u4,u5,u55,u6,u7}, and the fact that sQGP appears to exhibit near-conformal dynamics.  
Specifically, we concentrate on finite 't Hooft coupling corrections to two second order transport coefficients
of a CFT plasma  --- a relaxation time $\tau_\Pi$  and $\kappa$. These transport coefficients are the simplest to compute 
as they can be extracted from the equilibrium correlation functions\footnote{The coefficient $\lambda_1$ is necessary  to describe 
the Bjorken expansion \cite{bj} of plasma  and its finite coupling corrections will be discussed elsewhere.}.
While we perform the computations in the string theory dual to $\caln=4$ $SU(N_c)$ superconformal  Yang-Mills (SYM) theory,
following the arguments in \cite{u7}\footnote{A more detailed discussion is given in \cite{EtaS}.}, 
the results are expected to hold for any conformal gauge theory plasma in four dimensions,
and should provide a reasonable estimate for sQGP.

Following \cite{s3,m4}, the stress energy tensor $T^{\mu\nu}$ of a viscous conformal fluid to second order in derivatives of the 
four-velocity $u^{\mu}$ takes form
\begin{equation}
T^{\mu\nu}=\epsilon u^{\mu}u^{\nu}+P\left(g^{\mu\nu}+u^{\mu}u^{\nu}\right)+\Pi^{\mu\nu}\,,
\eqlabel{st1}
\end{equation}
where $\epsilon$, $P$ are the energy density and the pressure, and the  dissipative part $\Pi^{\mu\nu}$ is a sum of the 
first ($\Pi_1$) and the second order ($\Pi_2$) terms: 
\begin{equation}
\Pi^{\mu\nu}=\Pi_1^{\mu\nu}\left(\nabla_\alpha u_\beta; \{\eta\}\right)
+\Pi_2^{\mu\nu}\left(\nabla_\g\nabla_\alpha u_\beta; \{\eta.\tau_\Pi,\kappa,
\lambda_1,\lambda_2,\lambda_3\}\right)\,.
\eqlabel{defsigma}
\end{equation}
The four-velocity dependence of the dissipative part of the stress-energy tensor is uniquely fixed by conformal invariance 
up to a single phenomenological coefficient to the first order in the derivative expansion, and up to five additional 
coefficients to the second order in the derivative expansion. In the former case, a single coefficient is  a shear viscosity 
$\eta$, while the latter includes a relaxation time $\tau_\Pi$, three coefficients $\lambda_i$ (describing viscous terms bilinear in 
four-velocities), and  a $\kappa$-term (describing viscous hydrodynamics in curved backgrounds). In the linear regime,
and in Minkowski space-time,  \ie, setting $\lambda_i=\kappa=0$, the new viscous hydrodynamics of \cite{s3,m4} reduces 
to MIS theory \cite{m,is}. A crucial observation of \cite{s3,m4} was that the MIS regime of a strongly coupled four-dimensional 
conformal gauge theory plasma is simply inconsistent. In a specific example\footnote{This is in fact universal to all four dimensional 
CFT gauge theories 
with a string theory dual.} of $\caln=4$ SYM, and at infinite 't Hooft coupling $\lambda\equiv g_{YM}^2 N_c\to \infty$,
it was found that \cite{s0,s3,m4}
\begin{equation}
\begin{split}
&\frac{\eta}{s}=\frac{1}{4\pi}\,,\qquad \tau_\Pi =\frac{2-\ln 2}{2\pi T}\,, \qquad  \kappa=\frac{\eta}{\pi T}\,,\\
&\lambda_1=\frac{\eta}{2\pi T}\,,\qquad \lambda_2=-\frac{\eta}{\pi T}\,,\qquad \lambda_3=0\,,
\end{split}
\eqlabel{sugracoeff}
\end{equation} 
where $s$ is the entropy density, and $T$ is the temperature.

The finite coupling corrections were computed only for the shear viscosity. It was found in 
\cite{blo,bb,f3,bl} that
\begin{equation}
\frac{\eta}{s}=\frac{1}{4\pi}\left(1+\frac{120}{8}\zeta(3)\ \lambda^{-3/2}+\cdots\right)\,.
\eqlabel{esf}
\end{equation}

In this paper, extending analysis of \cite{blo,bb}, we find
\begin{equation}
\tau_\Pi T=\frac{2-\ln 2}{2\pi}+\frac{375}{32\pi}\zeta(3)\ \lambda^{-3/2}+\cdots\,,
\eqlabel{tfinite}
\end{equation}
\begin{equation}
\kappa=\frac{\eta}{\pi T}\left(1-\frac{145}{8}\zeta(3)\ \lambda^{-3/2}+\cdots\right)\,.
\eqlabel{kappafinite}
\end{equation}

The computations are quite technical, so we present only relevant steps and  for the details refer the reader to previous 
work on  the subject: \cite{ss,s3,blo,bb,bl}. In the next section we describe $\calo(\a'^{3})$ near-extremal D3 brane 
geometry \cite{d31,d32}, primarily to set-up our notation. In section 3, following \cite{blo,bl}, we compute $\{\tau_\Pi,\kappa\}$ 
from the retarded correlation function of the stress energy tensor.
In section 4, following \cite{bb,bl}, we compute the dispersion relation of 
a sound quasinormal mode up to third order in momentum, and confirm the value of $\tau_\Pi$ obtained in section 3.

\section{Background geometry}

The background black brane geometry dual to a strongly coupled $\caln=4$ SYM plasma at finite 't Hooft coupling was found in \cite{d31,d32}. 
The ten dimensional background takes the form 
\begin{equation}
\begin{split}
ds^2_{10} & = e^{-\frac { 10}{ 3}\nu} ds_5^2
 + e^{2 \nu} d\Omega_5^2\,, \\
F_5 & = \mathcal F_5+\star \mathcal F_5, \qquad \mathcal F_5=-4 dvol_{S_5}
\end{split}
\eqlabel{10m}
\end{equation}
where $d\Omega_5^2$ is a volume element of a round five-sphere, and 
\begin{equation}
ds_5^2\equiv g_{5\mu\nu}dx^\mu dx^\nu=\frac{r_0^2}{u} e^{c(u)} \left( - f e^{a(u)} dt^2 + 
 d\x^2 \right)  + \frac{du^2}{4 u^2 f} e^{b(u)}\,.
\eqlabel{mu5u}
\end{equation}
Here $f(u)=1-u^2$, $r_0$ is the parameter of non-extremality of the 
black brane geometry, and we set the ``AdS radius''
 $L$ to one. 
 To leading order in\footnote{The string tension $\alpha'$, or more precisely $\a'/L^2$, is identified with $\lambda^{-1/2}$ of the $\caln=4$ SYM.}  
\begin{equation}
\gamma\equiv \frac 18\zeta(3)\a'^3\,,
\eqlabel{defgamma}
\end{equation}
functions  $a$, $b$, $c$, $\nu$  were found in \cite{d31,d32}
\begin{equation}
\begin{split}
a(u) =& -15\, \gamma\, (5 u^2+5 u^4-3u^6)\,, \\
b(u) =& 15\, \gamma\, (5 u^2+5 u^4-19 u^6)\,,  \\
c(u) =& 0\,,  \\
\nu (u) =&\frac {15\gamma}{ 32} u^4 (1+u^2)\,.
\end{split}
\eqlabel{abcnu}
\end{equation}
The Hawking temperature corresponding to the metric \eqref{10m} is 
\begin{equation}
T \equiv T_0\left(1+15\g\right)= \frac{r_0}{ \pi} \left( 1 + 15 \gamma \right)\,.
\eqlabel{temperature}
\end{equation}
These corrections were found assuming the only relevant term at order $\gamma$ is $C^4$ \cite{C4}.
In \cite{F5} the full set of $\gamma$ corrections were computed including five-form terms, and there
it was found that the black D3-brane solution receives corrections only from the $C^4$ term. In \cite{EtaS}
it was further shown that the full spectrum of quasi-normal modes in this background is also unaffected by
the five-form terms, and therefore we are justified in this paper in working just with the $C^4$ correction.

\section{Relaxation time from the Kubo formula}

To obtain retarded correlation function of the boundary stress energy tensor, we study scalar perturbations of the 
background geometry \eqref{mu5u} (see \cite{ss,s3}): 
\begin{equation}
g_{5\mu\nu}\to g_{5\mu\nu}+h_{xy}(u,\x)\,.
\eqlabel{mfluc}
\end{equation}
It will be convenient to introduce a field  $\p (u,\x)$,
\begin{equation}
\p (u,\x) = \frac{u}{r_0^2} \, h_{xy} (u,\x)\,,
\eqlabel{defp}
\end{equation}
and use the Fourier decomposition
\begin{equation}
 \p (u,\x) = \int\! \frac{d^4 k}{(2\pi )^4} e^{-i\omega t + i \ka \cdot\x}
\varphi_k (u)\,.
\eqlabel{defpft}
\end{equation}
Finally, we introduce
\begin{equation}
\ww\equiv \frac{\omega}{2\pi T_0}\,,\qquad \ka\equiv \frac{k}{2\pi T_0}\,.
\eqlabel{hwdef0}
\end{equation}

\subsection{The effective action}
The effective action to order $\calo(\g)$ for $\p_k(u)$ takes form \cite{blo}:
\begin{equation}
\begin{split}
S_{eff} =& \frac{ N^2_c} {8 \pi^2 } \int  \frac{d^4 k} { (2 \pi)^4}
 \int_0^1 du \Biggl[
    \, A\,  \varphi_{k}''\varphi_{-k} +
 B\,  \varphi_{k}'\varphi_{-k}'  + 
 C \, \varphi_{k}'\varphi_{-k} 
 \\
 +& D\, \varphi_{k}\varphi_{-k} 
  + E\, \varphi_{k}''\varphi_{-k}'' 
+ F\,  \varphi_{k}''\varphi_{-k}'\Biggr]\,. 
\end{split}
\eqlabel{seff}
\end{equation}
The coefficients $A, B, C, D, E, F$ are even functions of the momentum.
They are given 
explicitly in appendix \ref{section_appendixa}.

Variation of $S_{eff}$ leads to 
\begin{equation}
\delta S_{eff} = \frac{N^2_c }{ 8\pi^2} \int  \frac{d^4 k }{ (2 \pi)^4}
\;  \Biggl[\;\;\;  \int_0^1 du \left( EOM \right) \delta \varphi_{-k}
+ \left( {\cal B}_1  \delta  \varphi_{-k} +
 {\cal B}_2  \delta  \varphi_{-k}'\right) \Biggl|_{0}^{1} \;\;\;   \Biggr]\,, 
\eqlabel{varact}
\end{equation}
where the coefficients of the boundary term are given by
\begin{equation}
 {\cal B}_1 = 
 - (A  \varphi_{k})' + 2 B  \varphi_{k}' + C  \varphi_{k} -
2 (E  \varphi_{k}'')' + F \varphi_{k}'' - (F  \varphi_{k}')'\,,
\eqlabel{boundary21}
\end{equation}
\begin{equation}
 {\cal B}_2 =  A  \varphi_{k} + F  \varphi_{k}' 
+ 2 E \varphi_{k}''\,,
\eqlabel{boundary2}
\end{equation}
and EOM denotes the left hand side of the 
Euler-Lagrange equation
\begin{equation}
A \varphi_{k}'' + C \varphi_{k}' + 2 D \varphi_{k} - \frac{d}{ d u}
\left( 2 B \varphi_{k}'  + C \varphi_{k} +  F \varphi_{k}''\right)
 +  \frac{d^2}{ d u^2}
\left( A \varphi_{k}  + 2 E \varphi_{k}'' +  F \varphi_{k}'\right) = 0 \,.
\eqlabel{ele}
\end{equation}

In order to have a well-defined variational principle, one 
has to add a generalized Gibbons-Hawking boundary 
term to the action  (\ref{seff}). As explained in \cite{blo},
this should be done perturbatively in $\g$. Specifically,
if we rewrite  \eqref{ele} in the form 
\begin{equation}
\varphi_{k}'' + p_1 \varphi_{k}' + p_0 \varphi_{k} = O(\gamma)\,, 
\eqlabel{eom}
\end{equation}
with all $\gamma$-dependent terms  exiled to the right,
the generalized  Gibbons-Hawking term $\calk_{gen}$ rendering  the variation of \eqref{varact}, takes form 
\begin{equation}
\calk_{gen} = -A\varphi_{k}\varphi_{-k}'-\frac F2 
\varphi_{k}'\varphi_{-k}' + E p_1 \varphi_{k}'\varphi_{-k}'
  + 2 E p_0 \varphi_{k}\varphi_{-k}'\,.
\eqlabel{gibbons}
\end{equation}
Notice that $\calk_{gen}$ differs from the standard (supergravity) Gibbons-Hawking term $\calk_{std}$:
\begin{equation}
\calk_{std}= -A\varphi_{k}\varphi_{-k}'+\calk_1 \varphi_{k}\varphi_{-k}'+\calk_2 \varphi_{k}\varphi_{-k}\,,
\eqlabel{ksrd}
\end{equation}
where the coefficients $\calk_1\propto \calo(\g)$ and $\calk_2\propto \calo(\g^0)$ are given in 
 appendix \ref{section_appendixa}. 

The bulk action \eqref{seff} can be rewritten in the form
 \begin{equation}
S_{eff} = \frac{N^2_c}{8\pi^2} \int  \frac{d^4 k}{(2 \pi)^4}
 \int_0^1 du \left( \partial_u {\cal B} + \frac 12
 \left[ EOM \right] \right)\,,
\eqlabel{seffb}
\end{equation}
where
\begin{equation}
\begin{split}
{\cal B} =& -\frac{A'}{ 2}  \varphi_{k} \varphi_{-k}  + 
 B  \varphi_{k}' \varphi_{-k}  +
\frac{C}{ 2}   \varphi_{k} \varphi_{-k}
-E'  \varphi_{k}'' \varphi_{-k}
+E  \varphi_{k}'' \varphi_{-k}'
- E  \varphi_{k}''' \varphi_{-k}
\\&+ \frac{F}{ 2}   \varphi_{k}' \varphi_{-k}'
- \frac{F'}{ 2}  \varphi_{k}' \varphi_{-k}\,.
\end{split}
\eqlabel{onshell}
\end{equation}
Thus on-shell, it reduces to the sum of  two boundary term: the horizon contribution ( as $u\to 1$ ) and the 
boundary contribution ( as $u\to 0$ ). In computing the two-point retarded correlation function of the 
boundary stress-energy tensor, the horizon contribution must be discarded \cite{ss1}; the boundary contribution 
is divergent as $u=\epsilon\to 0$ and must be supplemented by the counterterm action \cite{ps}:
\begin{equation}
\begin{split}
S_{ct}=&-\frac{3N_c^2}{4\pi^2}\int _{u=\epsilon} d^4x \sqrt{-\g}\left(1+\frac 12 P-\frac{1}{12}\left(P^{kl}P_{kl}-P^2\right)
\ln \epsilon\right)\,,
\end{split}
\eqlabel{sct}
\end{equation}
where $\g_{ij}$ is the metric induced at the $u=\epsilon$ boundary, and 
\begin{equation}
P=\g^{ij}P_{ij}\,,\qquad P_{ij}=\frac 12 \left(R_{ij}-\frac 16 R\g_{ij}\right)\,.
\eqlabel{defpp}
\end{equation}
We can further rewrite \eqref{sct} as\footnote{Note that the  conformal anomaly term 
in \eqref{sct} does not contributed to the retarded correlation functions at order $\calo(\ww^2,\ka^2)$.} 
\begin{equation}
\begin{split}
S_{ct}=-\frac{N^2_c}{8\pi^2} \int  \frac{d^4 k}{(2 \pi)^4} \biggl(\calt+\calo\left(\ww^4,\ww^2\ka^2,\ka^4\right)\biggr) 
\p_{k}\p_{-k}\,,
\end{split}
\eqlabel{sctft}
\end{equation}
where $\calt$ is given  in appendix \ref{section_appendixa}.

We would like to emphasize that the counterterm action \eqref{sct} is computed assuming the 
standard Gibbons-Hawking term \eqref{ksrd}. In particular, it takes into account  
the divergent as $u=\epsilon\to 0$ boundary 
contribution coming from ( the supergravity part of )  $\calk_2$ term in \eqref{ksrd}.  
Thus, even though at the level of the effective action \eqref{seff} the variational principle is well defined
just with $\calk_{gen}$,  in order to obtain finite retarded correlation functions with the counterterm action \eqref{sct},
we need to supplement $\calk_{gen}$ with the supergravity part\footnote{As observed in \cite{blo}, the $\calo(\g)$ 
parts of the coefficients in \eqref{gibbons}, \eqref{ksrd} (and also \eqref{sctft}) will not contribute 
in the limit $u=\epsilon\to 0$.} of $\calk_2 \p_k\p_{-k}$: 
\begin{equation}
\calk_{gen}\to \calk_{gen}+\calk_2 \p_k\p_{-k}\,.
\eqlabel{redefk}
\end{equation}
Altogether, the total renormalized boundary action takes the form
\begin{equation}
S_{tot}(\epsilon)=-\frac{N^2_c}{8\pi^2} \int  \frac{d^4 k}{(2 \pi)^4}\ \calf_{k}\bigg|_{u=\epsilon}\,,
\eqlabel{stot}
\end{equation}
where
\begin{equation}
\begin{split}
 {\cal F}_k =& \frac{N^2_c}{ 8 \pi^2} \Biggl[ \left( B - A\right) \varphi_{k}' \varphi_{-k} + \frac{1}{ 2}
 \left( C - A'+\calk_2+\calt\right) \varphi_{k} \varphi_{-k} -E'  \varphi_{k}'' \varphi_{-k}
+E  \varphi_{k}'' \varphi_{-k}' \\
&- E  \varphi_{k}''' \varphi_{-k} - \frac{F'}{ 2}  \varphi_{k}' \varphi_{-k} +
 E p_1 \varphi_{k}'\varphi_{-k}'
  + 2 E p_0 \varphi_{k}'\varphi_{-k}\Biggr]\,.
\end{split}
\eqlabel{boundaryterm}
\end{equation}
 
\subsection{The solution for $\p_k$}
We now turn to finding the solution to the equation of motion \eqref{ele}, or rather its equivalent to order $\calo(\g)$,
\eqref{eom}:
\begin{equation}
\begin{split}
&\p_k''-\frac{u^2+1}{u f} \p_k'+\frac{\ka^2 u^2+\ww^2-\ka^2}{uf^2}\p_k= -\frac 14\g \Biggl[\Biggl(3171 u^4+3840 \ka^2 u^3+2306 u^2\\
&-600\Biggr) u\  \p_k'
+ \frac{u}{f^2} \Biggl(600 \ww^2-300 \ka^2+50 u+(3456 \ka^2-2856 \ww^2) u^2+768 u^3 \ka^4\\
&+(-6560 \ka^2+2136 \ww^2) u^4+(-768 \ka^4-275) u^5+3404 \ka^2 u^6+225 u^7\Biggr)\ \p_k\Biggr]\,.
\end{split}
\eqlabel{sol1}
\end{equation}

The incoming wave boundary condition for $\p_{k}$ at the horizon implies \cite{bl} 
\begin{equation}
\p_k\propto (1-u^2)^{-\frac{i\ww(1-15\g)}{2}}\,,\qquad u\to 1_-\,.
\eqlabel{incoming}
\end{equation} 
Thus we represent the solution to \eqref{sol1} perturbatively in both $\g$ and $(\ww,\ka^2)$ as 
\begin{equation}
\begin{split}
\p_k=&(1-u^2)^{-\frac{i\ww(1-15\g)}{2}}\biggl(G_k^{(0)}+i \ww\ G_k^{(1)}+\ww^2\ G_k^{(2)} +\calo(\ww^3,\ww \ka^2)\biggr)\,,\\
G_k^{(i)}=&G_k^{(i,0)}+\g G_k^{(i,1)}+\calo(\g^2)\,,\qquad i=0,1,2\,,
\end{split}
\eqlabel{gi}
\end{equation}
where 
\begin{equation}
G_k^{(i)}(u)\equiv G_k^{(i)}(\ww,q^2;u)=\lambda^0\ G_k^{(i)}(\lambda\ww,\lambda^2\ka^2;u)\,,
\end{equation}
and 
\begin{equation}
\lim_{u\to 1_-}G_k^{(0)}(u)=1\,,\qquad \lim_{u\to 1_-}G_k^{(1)}(u)= \lim_{u\to 1_-}G_k^{(2)}(u)=0\,.
\eqlabel{ulim}
\end{equation}
Explicitly, we find:
\begin{equation}
\begin{split}
G_{k}^{(0,0)}=1\,,\qquad G_k^{(0,1)}=-\frac{25}{16} u^6-\frac{25}{16} u^4+\frac{25}{8}\,,
\end{split}
\eqlabel{order0}
\end{equation}
\begin{equation}
\begin{split}
G_{k}^{(1,0)}=0\,,\qquad G_{k}^{(1,1)}=\frac{43}{2} u^6+\frac{195}{2} u^2+\frac{135}{2} u^4-\frac{373}{2}\,,
\end{split}
\eqlabel{order1}
\end{equation}
\begin{equation}
\begin{split}
G_{k}^{(2,0)}=&\frac 14 \ln^2\left(\frac u2+\frac 12\right)+\ln\left(\frac u2+\frac 12\right)
-\frac 12 \dilog\left(\frac u2 +\frac 12\right)-\frac{\ka^2}{\ww^2} \ln\left(\frac u2+\frac 12\right)\,,
\\
G_{k}^{(2,1)}=&\left(\frac{25}{32} u^4+\frac{25}{32} u^6+\frac{215}{16}\right) \left(\dilog\left(\frac u2+\frac 12\right)
-\frac 12 \ln^2(1+u)\right)
+\biggl(\frac{215}{16} \ln 2+\frac{2445}{8}\\
&-\frac{195}{2} u^2+\left(\frac{25}{32} \ln 2-\frac{1105}{16}\right) u^4+\left(-\frac{369}{16}+\frac{25}{32} \ln 2\right) 
u^6\biggr) 
\ln(1+u)\\
&-\frac{2445}{8} \ln 2+\frac{2885}{6}-\frac{215}{32} \ln^2 2-\frac{605}{2} u+\left(\frac{195}{2} \ln 2-\frac{195}{2}\right) 
u^2-\frac{40}{3} u^3\\
&+\left(-\frac{135}{2}-\frac{25}{64} \ln^2 2+\frac{1105}{16} \ln 2\right) u^4
+\frac{43}{2} u^5\\
&+\left(-\frac{43}{2}-\frac{25}{64} \ln^2 2+\frac{369}{16} \ln 2\right) u^6
+\frac{\ka^2}{\ww^2} \Biggl(\left(\frac{1375}{8}+\frac{25}{16} u^6+\frac{25}{16} u^4\right) \ln(1+u)\\
&+\frac{821}{6}-\frac{1375}{8} \ln 2-175 u+\frac{195}{2} u^2-\frac{250}{3} u^3+\left(\frac{135}{2}-\frac{25}{16} \ln 2\right) 
u^4-65 u^5\\
&+\left(\frac{43}{2}-\frac{25}{16} \ln 2\right) u^6\Biggr)\,.
\end{split}
\eqlabel{order2}
\end{equation}

\subsection{Coupling constant correction to relaxation time}
Having found the solution for a gravitational perturbation, 
we can compute the correlation function
 $G_{xy,xy}(\omega,q)$ by applying the  Minkowski AdS/CFT prescription
\cite{ss1}
 \begin{equation}
 G_{xy,xy}^R(\omega,q) = \lim_{u\rightarrow 0}  \frac{2 {\cal F}_q}{|\p_q|^2}\,.
\label{prescription}
\end{equation}

Explicitly we find
 \begin{equation}
\begin{split} 
G_{xy,xy}^R(\omega,q) =& \frac{ \pi^2 N^2_c T^4(1+15\g)}{ 4}
\Biggl(\frac 12-i\hw \biggl[1+120 \g\biggr]+\biggl[
-\hq^2+\hw^2-\hw^2\ln 2 \\
&+\g\left(-120 \hw^2\ln 2 +25 \hq^2+\frac{905}{2} \hw^2\right)\biggr]+\calo(\hw^3,\hw \hq^2)\Biggr)+\calo(\g^2)\,,
\end{split}
\eqlabel{limit}
\end{equation}
where we used \eqref{temperature} to reintroduce the temperature, and denoted
\begin{equation}
\hw\equiv \frac{\omega}{2\pi T}\,,\qquad \hq\equiv \frac{q}{2\pi T}\,.
\eqlabel{hwdef}
\end{equation}

In the hydrodynamic limit the retarded correlation function $G_{xy,xy}^R(\omega,q)$ takes form \cite{s3}
\begin{equation}
G_{xy,xy}^R(\omega,q) =P-i\eta \omega+\eta\tau_\Pi\omega^2-\frac{\kappa}{2}\left(\omega^2+q^2\right)
+\calo(\omega^3,\omega q^2)\,.
\eqlabel{hydro}
\end{equation}
Comparing \eqref{limit} and \eqref{hydro} we conclude
\begin{equation}
\begin{split}
&P=\frac{\pi^2N_c^2T^4}{8}\biggl(1+15\g+\calo(\g^2)\biggr)\,,\qquad \frac{\eta}{s}=\frac{1}{4\pi}\biggl(1+120\g
+\calo(\g^2)\biggr)\,,\\
&\tau_\Pi T=\frac{2-\ln 2}{2\pi}+\frac{375}{4\pi}\g+\calo(\g^2)\,,\qquad \kappa=\frac{\eta}{\pi T}\biggl(1-145\g
+\calo(\g^2)\biggr)\,.
\end{split}
\eqlabel{main}
\end{equation}
Eq.~\eqref{main} is our main result.

\section{Relaxation time from the sound pole}

The equation of motion for the sound quasinormal mode has been obtained in \cite{bb}.
We represent it here in an equivalent form, to order $\calo(\g)$:
\begin{equation}
\begin{split}
0=&Z''_{sound}+\frac{3 \ww^2+(3 x^2-2) \qq^2}{x (3 \ww^2-(x^2+2) \qq^2)} Z'_{sound}
\\
&+\frac{3 \ww^4-(4 x^2 (1-x^2)^{3/2}+ 2\ww^2 (2x^2+1)) \qq^2+x^2 (x^2+2) \qq^4}{(3 \ww^2-(x^2+2) \qq^2) x^2 (1-x^2)^{3/2}} Z_{sound}
\\
&+J_{sound}[Z_{sound}]+\calo(\g^2)\,,
\end{split}
\eqlabel{zsound}
\end{equation}
where $Z_{sound}=Z_{sound}(x)$, $x=\sqrt{1-u^2}$ and $(\ww,\qq)$ are the dimensionless momenta reduced with respect to $2\pi T_0$, \eqref{hwdef0}.
The order $\calo(\g)$ the  source term $J_{sound}$ is given by
\begin{equation}
\begin{split}
J_{sound}=& -\frac{\g}{4 x^2 \biggl(3 \ww^2-(x^2+2) \qq^2\biggr)^3 (1-x^2)^{1/2}} \Biggl[ Z_{sound}'\ x^3\ J_{sound,0} - Z_{sound}\ 
J_{sound,1}\Biggr]\,,
\end{split}
\eqlabel{ssource}
\end{equation} 
\begin{equation}
\begin{split}
J_{sound,0}=&256 \qq^2 (1-x^2)^2 (3 \ww^2-(x^2+2) \qq^2) \biggl(15 \ww^4-30 \ww^2 (5 x^2+2) \qq^2\\
&+(116 x^2+44+35 x^4) \qq^4\biggr)+\biggl(27 \ww^6 (3171 x^4+4877-8648 x^2)\\
&+27 \ww^4 (20291 x^2-12394-9790 x^4+3293 x^6) \qq^2-9 \ww^2 (-24153 x^4+54172 x^2\\
&-31188+51661 x^8-47492 x^6) \qq^4+(-322145 x^6+220694 x^8-77416\\
&+89709 x^{10}-52806 x^4+147364 x^2) \qq^6\biggr) (1-x^2)^{1/2}\,,
\end{split}
\eqlabel{ssource0}
\end{equation} 
\begin{equation}
\begin{split}
J_{sound,1}=&648 \ww^8 (-5-59 x^2+89 x^4)-36 \ww^6 (-3989 x^4-180+7003 x^6-1259 x^2) \qq^2\\
&+12 \ww^4 \qq^4 (-40299 x^4+6242 x^2-360+159 x^6+40333 x^8)-12 \ww^2 \qq^6 (37927 x^8\\
&-80+8900 x^2-30499 x^6+17099 x^10-29972 x^4)+4 x^2 (x^2+2) (5811 x^8\\
&+18043 x^6-14991 x^4-12192 x^2+4004) \qq^8+\biggl(-675 x^2 \ww^6 (9 x^4-16 x^2+5)\\
&+9 x^2 \ww^4 \qq^2 (8190+87357 x^6+12989 x^2-111386 x^4)+\qq^4 ((-79980 x^8\\
&-820917 x^{10}+1316595 x^6-284748 x^4-83700 x^2) \ww^2+20736 x^2 (1-x^2) \ww^6)\\
&+\qq^6 (x^2 (131316 x^2+134283 x^{10}+24040+410574 x^8-430386 x^4-292777 x^6)\\
&+(20736 x^6+20736 x^4-41472 x^2) \ww^4)+(27648 \ww^2 x^2-20736 x^6 \ww^2\\
&-6912 x^8 \ww^2) \qq^8-768 \qq^{10} x^2 (1-x^2) (x^2+2)^3\biggr)
(1-x^2)^{1/2}\,.
\end{split}
\eqlabel{ssource1}
\end{equation} 
The incoming wave boundary condition at the horizon implies the following perturbative expansion in the hydrodynamic limit 
\cite{bl}
\begin{equation}
\begin{split}
Z_{sound}=&x^{-i\ww(1-15\g)}\biggl(z_{sound}^{(0)}+i \qq z_{sound}^{(1)}+\qq^2 z_{sound}^{(2)}+\calo(\ww^3,\ww \qq^2)\biggr)\,,\\
z_{sound}^{(i)}=&z_{sound,0}^{(i)}+\g z_{sound,1}^{(i)}+\calo(\g^2)\,,
\end{split}
\eqlabel{zso}
\end{equation}
where 
\begin{equation}
z_{sound}^{(i)}(x)\equiv z_{sound}^{(i)}(\ww,\qq^2;x)=\lambda^0\ z_{sound}^{(i)}(\lambda\ww,\lambda^2\qq^2;x)\,,
\end{equation}
and 
\begin{equation}
\lim_{x\to 0_+}z_{sound}^{(0)}(x)=1\,,\qquad \lim_{x\to 0_+}z_{sound}^{(1)}(x)=\lim_{x\to 0_+}z_{sound}^{(2)}(x)=0\,.
\eqlabel{xlimit}
\end{equation}
The dispersion relation for the sound quasinormal modes is obtained by imposing a Dirichlet condition on $Z_{sound}$ at the boundary:
\begin{equation}
\lim_{x\to 1_-}Z_{sound}(x)=0\qquad \Longleftrightarrow\qquad \ww\equiv \ww(\qq)\,.
\eqlabel{disp}
\end{equation}

\subsection{Sound quasinormal spectrum to $\calo(\qq^2)$}

The sound quasinormal mode and and its spectrum to the leading and the first subleading order in the 
hydrodynamic approximation was found in \cite{bb,bl}:
\begin{equation}
\begin{split}
z_{sound,0}^{(0)}=\frac{3\ww^2+(x^2-2)\qq^2}{3\ww^2-2\qq^2}\,,\qquad z_{sound,0}^{(1)}=\frac {2\ww\qq x^2}
{3\ww^2-2\qq^2}\,,
\end{split}
\eqlabel{ssss3}
\end{equation}
\begin{equation}
\begin{split}
z_{sound,1}^{(0)}=&\frac{5x^2}{16(3\ww^2-2\qq^2)^2}\biggl(\qq^4\left(2404+446x^2-4164x^4+2006x^6\right)
\\
&-3\ww^2\qq^2\left(1588+183x^2-2072x^4+1003x^6\right)+45\ww^4\left(5-4x^2+x^4\right)
\biggr)\,,
\\
z_{sound,1}^{(1)}=&\frac{\ww x^2}{8\qq (3\ww^2-2\qq^2)^2}\biggl(\qq^4\left(-13344+5846x^2-4520x^4+1734x^6\right)
\\
&-3\ww^2\qq^2\left(-9744+5035x^2-2604x^4+867x^6\right)\\
&-36\ww^4\left(594-264x^2+43x^4\right)
\biggr)+\frac{30\ww\qq x^2}{2\qq^2-3\ww^2}
\end{split}
\eqlabel{ssss4}
\end{equation}
\begin{equation}
\begin{split}
\ww(\qq)=\frac{1}{\sqrt{3}}\qq-i\qq^2\left(\frac 13+\frac{105}{3}\g\right)+\calo(\qq^3,\g^2)\,. \\
\end{split}
\eqlabel{results}
\end{equation}

\subsection{Sound quasinormal spectrum to $\calo(\qq^3)$}
Here, we extend the analysis of the previous section to the next order in the hydrodynamic approximation.
Since our ultimate goal is to determine the dispersion relation \eqref{disp} to order $\calo(\qq^3)$, it is sufficient 
to use the zeroth order dispersion relation, \ie, to set $\ww=\frac{1}{\sqrt{3}}\qq$. This drastically simplifies the 
hydrodynamic equations for $z_{sound,0}^{(2)}$ and $z_{sound,1}^{(2)}$. We find it more convenient to solve the 
resulting equations using the $u=\sqrt{1-x^2}$ variable.

Explicitly we find 
\begin{equation}
\begin{split}
z_{sound,0}^{(2)}(u)=&\frac{u^2}{6} \left(\frac 12 \ln^2\left(\frac u2+\frac 12\right)-\dilog\left(\frac u2+\frac 12\right)\right)
+\frac 23 \ln\left(\frac u2+\frac 12\right)\\
&+\frac 23 (u+2) (1-u)\,,
\end{split}
\eqlabel{z20}
\end{equation}
\begin{equation}
\begin{split}
z_{sound,1}^{(2)}(u)=&u^2 \calc_2+\left(2+\frac {u^2}{2} \ln(1-u^2) \right) \calc_1+\frac{u^2}{96} \cali_1(u)
\\
&-\frac{1}{24}  \left(1+\frac{u^2}{4} \ln(1-u^2) \right)\cali_2(u)\,,
\end{split}
\eqlabel{z21}
\end{equation}
where 
\begin{equation}
\begin{split}
\cali_1(u)=&-\int_0^u dt\ \frac{4+t^2\ln(1-t^2)}{(2-t^2)^3 (t+1)}\times \cali(t)\,,
\end{split}
\eqlabel{cali1}
\end{equation}
\begin{equation}
\begin{split}
\cali_2(u)=&\int_0^u dt \ \frac{t^2}{(2-t^2)^3 (t+1)}\times \cali(t)\,,
\end{split}
\eqlabel{cali2}
\end{equation}
\begin{equation}
\begin{split}
\cali(t)=&\Biggl\{10 t^3 (t+1) (9027 t^6-43808 t^4+58722 t^2\\
&-23100) \left(\dilog\left(\frac t2+\frac12\right)-\frac 12 \ln^2(1+t)\right)+ 2 (t+1) t (10000+(376440\\
&-115500 \ln 2) t^2+(293610 \ln 2-1104128) t^4+(-219040 \ln 2+847972) t^6\\
&+(45135 \ln 2-179418) t^8) \ln(1+t)-18440+(-20000 \ln 2-12568) t\\
&+(-20000 \ln 2+1912) t^2+(543780-752880 \ln 2+115500 \ln^2 2) t^3\\
&+(1143490+115500 \ln^2 2-752880 \ln 2) t^4+(-293610 \ln^2 2-1403332\\
&+2208256 \ln 2) t^5+(-2629768+2208256 \ln 2-293610 \ln^2 2) t^6\\
&+(219040 \ln^2 2-1695944 \ln 2+946434) t^7+(1729061+219040 \ln^2 2\\
&-1695944 \ln 2) t^8+(-45135 \ln^2 2+358836 \ln 2-188378) t^9+(-340799\\
&+358836 \ln 2-45135 \ln^2 2) t^{10}\Biggr\}\,,
\end{split}
\eqlabel{cali}
\end{equation}
and the integration constants $\calc_i$ are tuned to satisfy the horizon boundary condition:
\begin{equation}
\lim_{u\to 1_-}z_{sound,1}^{(2)}(u)=0\,.
\eqlabel{hbc2}
\end{equation}
The latter is achieved provided 
\begin{equation}
\calc_1=\frac{1}{48}\cali_2(1)\,,\qquad \calc_2=-\frac{1}{96}\cali_1(1)\,.
\eqlabel{solvec1c2}
\end{equation}

If we denote
\begin{equation}
\lim_{u\to 0_+} z_{sound,1}^{(2)}(u)=2\calc_1\equiv z_{1,0}^{(2)}\,,
\eqlabel{z0}
\end{equation}
the Dirichlet boundary condition \eqref{disp} will lead to the following dispersion 
relation for the sound quasinormal mode
\begin{equation}
\begin{split}
\ww(\qq)=&\frac{1}{\sqrt{3}}\qq-i\qq^2\left(\frac 13+\frac{105}{3}\g\right)
+\qq^3\Biggl(\frac{3- 2 \ln 2}{6\sqrt{3}}\\
&+\frac{1}{24\sqrt{3}} \biggl(-2758+12 z_{1,0}^{(2)}+1705 \ln 2\biggr)\g\Biggr)+\calo(\qq^4,\g^2)\,. \\
\end{split}
\eqlabel{results1}
\end{equation}

We were unable to evaluate \eqref{solvec1c2} analytically; numerically, we find
\begin{equation}
z_{1,0}^{(2)}=264.7598406\,.
\eqlabel{numz21h}
\end{equation}

\subsection{Relaxation time from the sound quasinormal spectrum}
Second order relativistic hydrodynamics of conformal fluids implies the following dispersion relation for the sound mode \cite{s3}
\begin{equation}
\omega=c_s q-i\Gamma q^2+\frac{\Gamma}{c_s}\left(c_s^2\tau_\Pi-\frac{\Gamma}{2}\right)k^3+\calo(k^4)\,,
\eqlabel{soundd}
\end{equation}
where 
\begin{equation}
\Gamma=\frac {2\eta}{3 sT}\,.
\eqlabel{gammadef}
\end{equation}
Comparing \eqref{results1} and \eqref{soundd} we find
\begin{equation}
c_s=\frac{1}{\sqrt{3}}+0\cdot \g+\calo(\g^2)\,,\qquad \Gamma T=\frac{1}{6\pi}\biggl(1+120\g\biggr)+\calo(\g^2)\,,
\end{equation}
in agreement with the conformal equation of state at order $\calo(\g)$, as well as in agreement with the ratio 
$\frac {\eta}{s}$ as given by \eqref{main}.
Additionally, we compute 
\begin{equation}
\tau_\Pi T=\frac{2-\ln 2}{2\pi}+\frac{1}{16\pi}\biggl(2425 \ln 2-3358+12 z_{1,0}^{(2)}\biggr)\g+\calo(\g^2)\,.
\eqlabel{tauTs}
\end{equation}
A required agreement between \eqref{main} and \eqref{tauTs} provides a prediction for $z_{1,0}^{(2)}$
\begin{equation}
z_{1,0}^{(2)}\bigg|_{prediction}=\frac{2429}{6}-\frac{2425}{12}\ln 2\,,
\eqlabel{predic}
\end{equation}
which is in excellent agreement with the actually numerical result \eqref{numz21h}.
Thus we have a highly nontrivial check on our analysis.

\section*{Acknowledgments}
We would like to thank Rob Myers, Aninda Sinha and Sam Vazquez for valuable discussions.
AB research at Perimeter Institute is supported in part by the Government
of Canada through NSERC and by the Province of Ontario through MRI.
AB gratefully acknowledges further support by an NSERC Discovery
grant and support through the Early Researcher Award program by the
Province of Ontario. MP work is supported by the Portuguese Fundacao para a Ciencia e Tecnologia, grant SFRH/BD/23438/2005.
MP also gratefully acknowledges Perimeter Institute for its hospitality.

\appendix

\section{Coefficients of the effective action, the Gibbons-Hawking term, and the counterterm}
\label{section_appendixa}

\begin{equation}
\begin{split}
A =&\frac{r_0^4}{ 2 u} \Biggl[ 8 f(u) + \gamma u^2 \Biggl(-600+25 u^2+(-44 \ww^2+44 \ka^2) u^3+1760 u^4-44 \ka^2 u^5\\
&-1185 u^6\Biggr)\Biggr]\,. 
\end{split}
\eqlabel{ca1}
\end{equation}
\begin{equation}
\begin{split}
B =&\frac{r_0^4}{ 8 u}  \Biggl[ 24 f(u)  + \gamma u^2 \Biggl(
-1800+179 u^2-(768 \ka^2+768 \ww^2) u^3+5424 u^4+768 \ka^2 u^5\\
&-3387 u^6) \Biggr)\Biggr]\,. 
\end{split}
\eqlabel{ca2}
\end{equation}
\begin{equation}
\begin{split}
C =&  - \frac{r_0^4}{ 4 u^2 f} \Biggl[ 8 f(u) (3+u^2) 
+   \gamma u^2 \Biggl(600-825 u^2+(-104 \ka^2+104 \ww^2) u^3-16945 u^4\\
&+(384 \ka^2+872 \ww^2) u^5+34005 u^6-280 \ka^2 u^7-16835 u^8\Biggr)\Biggr]\,.
\end{split}
\eqlabel{ca3}
\end{equation}
\begin{equation}
\begin{split}
D =&   \frac{r_0^4}{ 8 u^3 f^2} \Biggl[ 16 f(u)^2 + 8 u f(u) \ww^2 
 +  \gamma u^3
 \Biggl(600 \ww^2+250 u+(-25 \ka^2+25 \ww^2) u^2\\
&+(944 \ka^2 \ww^2+296 \ww^4+296 \ka^4-2120) u^3+(-2400 \ww^2+825 \ka^2) u^4+(1570\\
&-592 \ka^4-944 \ka^2 \ww^2) u^5+(1007 \ww^2-1575 \ka^2) u^6+(296 \ka^4+2220) u^7+775 \ka^2 u^8\\
&-1920 u^9\Biggr) \Biggr]\,.
\end{split}
\eqlabel{ca4}
\end{equation}
\begin{equation}
\begin{split}
E =&   \gamma \, 37\,  r_0^4  u^5 f(u)^2\,. 
\end{split}
\eqlabel{ca5}
\end{equation}
\begin{equation}
\begin{split}
F =&  \gamma\,  2 r_0^4 u^4 f(u) \, (11-37 u^2)\,. 
\end{split}
\eqlabel{ca6}
\end{equation}
\begin{equation}
\begin{split}
\calk_1 =& -\g\ r_0^4 u^3 \Biggl(-25+(44 \ww^2-44 \ka^2) u+160 u^2+44 \ka^2 u^3-135 u^4\Biggr)\,.
\end{split}
\eqlabel{ca7}
\end{equation}
\begin{equation}
\begin{split}
\calk_2 =&-\frac{4 r_0^4}{u^2} \Biggl[u^2-2+15\g u^2 \Biggl(2 u^6-8 u^4+5\Biggr)\Biggr]\,.
\end{split}
\eqlabel{ca8}
\end{equation}
\begin{equation}
\begin{split}
\calt =&-\frac{r_0^4}{2u^2\sqrt{1-u^2}} \Biggl[6+(-2 \ww^2+2 \ka^2) u-6 u^2-2 \ka^2 u^3\\
&+15 \g u^2 \biggl(3 u^4-5 u^2-5\Biggr) 
\Biggl(3+(\ww^2+\ka^2) u-3 u^2-\ka^2 u^3\Biggr) \Biggr]\,.
\end{split}
\eqlabel{ca9}
\end{equation}

\end{document}